\documentclass[format=acmsmall,screen=true]{acmart} 

\usepackage{booktabs} 
\usepackage{longtable}

\usepackage{pdflscape} 

\usepackage{url}

\usepackage{array}
\newcolumntype{P}[1]{>{\centering\arraybackslash}p{#1}}

\begin{document}

		\title[Performance-Aware Management of Cloud Resources: A Taxonomy and Future Directions]{Performance-Aware Management of Cloud Resources: A Taxonomy and Future Directions} 
		
		\author{Sara Kardani Moghaddam}	
		\affiliation{%
			\institution{School of Computing and Information Systems, The University of Melbourne}		
			\state{VIC}
			\postcode{3010}
			\country{Australia}}
		
		\author{Rajkumar Buyya}		
		\affiliation{%
				\institution{School of Computing and Information Systems, The University of Melbourne}
				\state{VIC}
			\postcode{3010}
				\country{Australia}}
		
		\author{Ramamohanarao Kotagiri}		
		\affiliation{%
			\institution{School of Computing and Information Systems, The University of Melbourne}
			\state{VIC}
		\postcode{3010}
			\country{Australia}}

\begin{abstract}
Dynamic nature of cloud environment has made distributed resource management process a challenge for cloud service providers. The importance of maintaining the quality of service in accordance with customer expectations as well as the highly dynamic nature of cloud hosted applications add new levels of complexity to the process. Advances to the big data learning approaches have shifted conventional static capacity planning solutions to complex performance-aware resource management methods. It is shown that the process of decision making for resource adjustment is closely related to the behaviour of the system including the utilization of resources and application components. Therefore, a continuous monitoring of system attributes and performance metrics provide the raw data for the analysis of problems affecting the performance of the application. Data analytic methods such as statistical and machine learning approaches offer the required concepts, models and tools to dig into the data, find general rules, patterns and characteristics that define the functionality of the system. Obtained knowledge form the data analysis process helps to find about the changes in the workloads, faulty components or problems that can cause system performance to degrade. A timely reaction to performance degradations can avoid violations of the service level agreements by performing proper corrective actions including auto-scaling or other resource adjustment solutions. In this paper, we investigate the main requirements and limitations in cloud resource management including a study of the approaches in workload and anomaly analysis in the context of the performance management in cloud. A taxonomy of the works on this problem is presented which identifies main approaches in existing researches from data analysis side to resource adjustment techniques. Finally, considering the observed gaps in the general direction of the reviewed works, a list of the new research gaps for future researchers is proposed.
\end{abstract}

\keywords{Anomaly Detection, Performance Management, Resource Management, Big-Data Analytics}

\maketitle

\section{Introduction}
Cloud computing as an on-demand, pay-as-you-go environment has been modelled based on two main concepts of elasticity and virtualization. The inherent flexibility brought by these techniques in the area of high performance computing is accompanied with the complexity of managing distributed resources while meeting the expectations of the users. The emergence of the public Cloud Service Providers (CSP) such as Amazon and Google which are extending the scientific limited applications of the cloud environment to industrial, academic and personal use cases make the need for more advanced and complex resource management solutions highly important. 

The main goal for CSPs is to find better ways of utilizing resources while keeping the service level agreements (SLA) as expected. SLAs are contracts among CSPs and customers to maintain the minimum Quality of Service (QoS) delivered by the offered applications. Breach of the SLAs costs the CSPs both money and their reputation. Considering dynamic characteristics of cloud including unreliability and heterogeneity in resources and workloads, simple static resource planning solutions does not work. Therefore, traditional resource management infrastructure is extended with monitoring modules which can provide timely information on the performance of the application along with the resource utilization of system components. The collected data from monitoring the system and application provide a source of highly valuable information about the health of the system. On the other hand, advances in data learning methods have provided missing parts of a data aware performance management offering all the concepts and tools for analyzing data to find patterns, trends and interesting changes in the behaviour of monitored components. The integration of two parts of performance data analytics and automated resource management brings new challenges and opportunities in both areas of theoretical concepts and practical implementations. In this paper, we try to identify the major challenges and corresponding solutions in the problem of data aware performance analysis and resource management in cloud. We present a taxonomy to depict various perspectives of the performance management in cloud, covering all aspects of data collection, analytics, and resource adjustment solutions.  

\subsection{Related Surveys and Our Contribution}
Although there are a number of survey and review articles identifying various aspects of data analysis or cloud resource management, they are more focused on one side of the problem without considering the requirements of other parts of data oriented performance management frameworks. For example, \citeauthor{Chandola:2009:ADS} present a survey discussing the general concept of abnormality in the data including various types of anomaly and applications of anomaly detection in the context of different problems. The paper presents a high level review of specific data requirements as well as mathematical models and algorithms such as classification and clustering methods to extract hidden information on existing patterns or features of the data. \citeauthor{Ibidunmoye:2015:PAD} investigate the anomaly problems in specific areas of performance analysis and bottleneck identification in computing systems and applications. They present various factors contributing to the performance anomaly problems including the types of bottlenecks, the granularity of knowledge expected from data analysis and possible algorithms to solve these problems. On the other hand, \citeauthor{Qu:2016:ASWTS} present a taxonomy on the resource scaling problem, focusing on the challenges of distributed resource management in the context of large web applications hosted on cloud platforms. They identify the challenges of dynamic resource management to meet specific requirements of web applications and categorize various scaling solutions to manage resource requirements of the application. 

 In contrast to these works, our work has a more integrated view on the problem of performance-aware resource management in the cloud and covers both areas application dependent workload analysis and anomaly detection techniques and their contribution in the resource management and particularly auto-scaling methods as the main resource level solutions for cloud hosted applications. We have also tried to specifically cover the works that integrate both sides of performance data analysis and corresponding resource adjustment techniques, implementing the complete circle of performance monitoring and data collection, data analysis, planning and decision making, and the execution of selected actions. Moreover, we have a more updated review of the recent works in the area as well as new discussions on the research gaps and directions for future researchers. We also present a taxonomy of source of performance related problems, data analysis methods as well as strategies to detect and handle anomalies including scaling techniques. 

\subsection{Paper Organization}
The rest of the paper is organized as follow: Section \ref{background} describes the main blocks of data aware resource management and existing challenges followed by listing the most influencing factors in this area. Section \ref{BDPM} introduces two main approaches in utilizing data as a source of extra knowledge for resource management. Then, sections \ref{Data_Level} to \ref{actions} reviews different characteristics of data analysis and resource management modules based on the categories identified in the taxonomy. Finally, section \ref{Gaps} discusses the main gaps and directions for future researchers and section \ref{Summary} concludes the paper.

\begin{figure}
	\centering
	\includegraphics[scale=0.59]{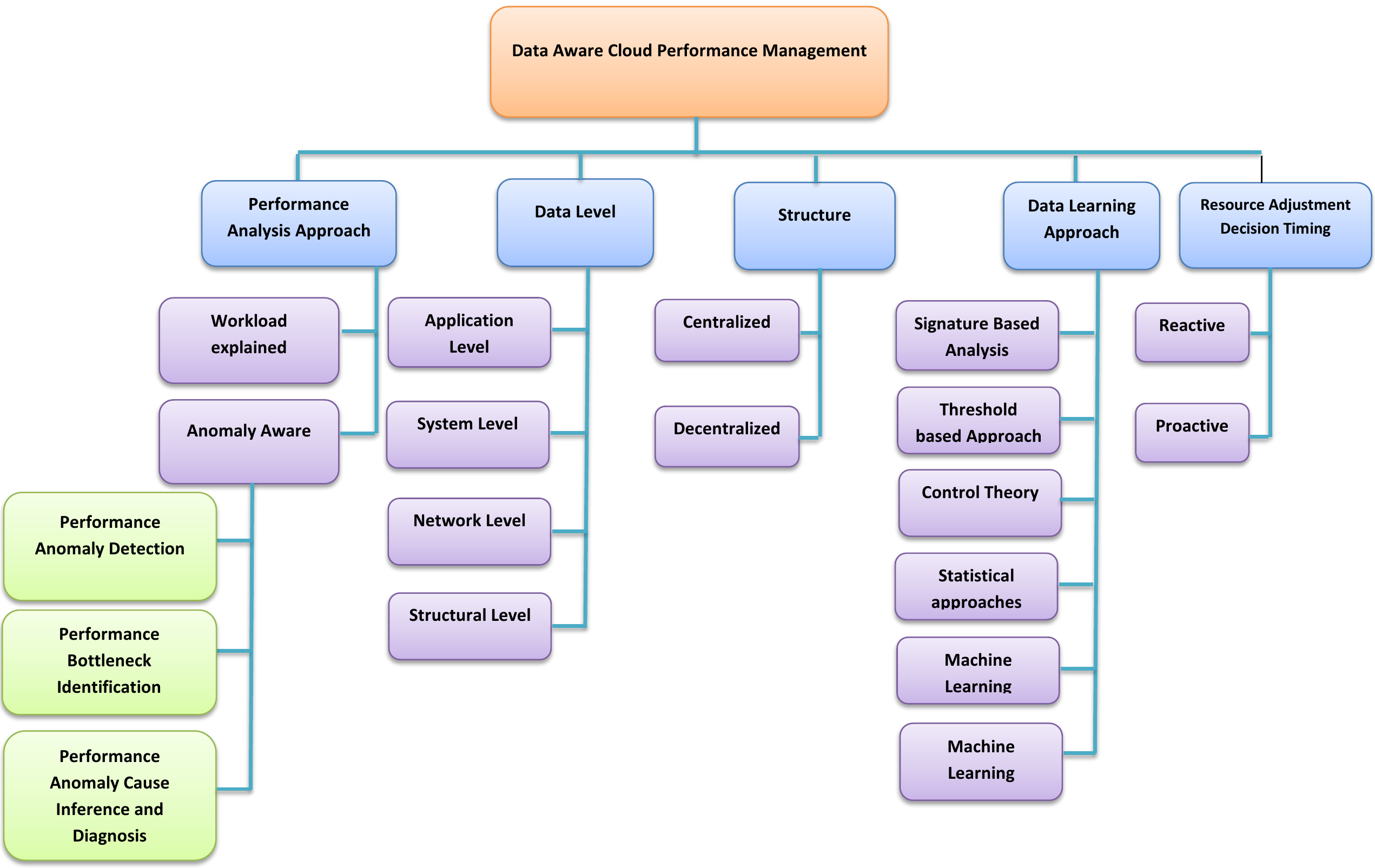}
	
	\caption{\label{fig:tax}
		The Taxonomy of Data-Aware Performance Management in Clouds.}
\end{figure}

\section{Background}\label{background}
The concept of resource management in cloud environment encompasses all the techniques and procedures that help to adjust the configuration of resources according to the demands of the users and applications in the system. For example, auto-scaling solutions are based on a characteristic that allows system resources expand or shrink at different levels of granularity (Virtual Machines (VMs), CPU, RAM, etc) automatically according to the perceived state of the system. To be clear about these concepts, we pursue following definitions in the rest of the paper:
	
	\textbf{\textit{Performance Indicators}}: All the measurable attributes from the resources and applications which demonstrate the degree of functionality of the corresponding unit in the system. These indicators continuously change over time and are initial sources of information for the health of the system. For example, the time it takes to load a page known as \textit{response time} (RT) from a web based application is the most perceptible sign that the system is performing at the expected level or not. Longer than usual RTs trigger the warnings of having some sort of the problem, requiring technical considerations form the system administrators.

	\textbf{\textit{System State}}: State or behaviour of the system at each time is an abstract representation of all the operational attributes and performance indicators of the system which can be recognized in normal or abnormal/anomalous condition. 
	
	The main indicators of an abnormal state are the presence of unexpected pattern or values in the performance indicators of the system.
	
	\textbf{\textit{Performance Degradations}} are caused by the abnormal behaviors when they affect the performance indicators adversary. For example, in the case of the increase in the number of requests (increased demand from customers) to a web server, if current resources cannot handle the newly received requests, the RT observed by users will increase. The unacceptable increases in the RT are considered as performance degradation which should be avoided. One solution can be to add new resources corresponding to the overloaded component of the application so the amount of resources is in accordance with the incoming load to the system. 
 
 Considering aforementioned definitions, any automated Resource Management Module (RMM) is dealing with two main challenges:

\emph{When a performance degradation is happening in the system?}
In an ideal, highly reliable environment where no abnormal behaviour is expected and applications show consistent behaviour with a stable performance, traditional static scheduling solutions will work and dynamic scaling of the resources is not required. However, in a real environment with a wide range of internal and external factors which can affect the behaviour of the system, performance degradation has become an important challenge to be dealt with accurately. There is a wide range of causes identified for these problems from fluctuations in the incoming workload to a malfunctioned hardware or a buggy software that can affect the performance of the application or VMs. Therefore, the onset time of the degradations should be known so a proper and timely corrective action can be started. Monitoring sensors which follow the performance of each component generate vast amount of data which include hidden patterns and signs of the health of the system. Previously, we had to rely on the expert of the human operators to skim the data and find alerting behavior. However, considering the scale of the generated data from hundreds and thousands of machines located in different geographical locations, the manual approach is not feasible anymore. Therefore, researchers have started to take advantage of the advanced data analytics methods and more powerful and cost-effective computing hardware to automate and accelerate the process and find better quality knowledge about the performance of the target systems.

\emph{What type of corrective action should be performed?}  
In order to alleviate the performance problems of the system, RMM should start a corrective action in the form of load redistribution, resource provisioning, migrations, etc. 
Current resource providers such as Amazon or Azure offer migrations or simple threshold based scaling services which change the number of VMs in the system. There are also more customized resource management policies such as on the fly changes in the resource configuration of one VM which is offered by some CSPs such as \cite{Profitbricks}. The selection of proper action can be dependent on many factors including technical or business limitations, type of the problem, etc. We have identified some of the most important factors as follows:
\begin{itemize}
	
	\item \textbf{Technical limitations:} Virtualization is the key concept for cloud models. It enables hosting different applications or the components of one application independently on one Physical Machine (PM) with migrations option available to move them to other PMs without significant down times in the system. Currently, many public resource providers such as Amazon and Microsoft Azure offer the required environment for CSPs to host their application on VMs and dynamically add/remove VMs in the system. There are also more fine grained controls available to configure resources at VMs level defined as vertical scaling. In this process, size of the VM can change on-the-fly without any rebooting of the VMs. However, the functionality needs support form both hypervisor and the kernel of the VM. Currently, providers such as Amazon \cite{amazon:2018, azure:2018} and Microsoft Azure do not support this functionality. 
	\item \textbf{Business considerations:} There are a vast amount of the resources offered by cloud resource providers with various pricing strategies. For example, Amazon offers on-demand instances with hour/seconds based pricing or much cheaper reserved instances with long-term contracts \cite{amazon:2018}. There are different pricing rules for vertical scaling of the VMs such as the offered rules by \cite{Profitbricks}. CSPs should consider these options when deciding on the configuration of their system and scaling policies. As a result, selecting the best action will be limited by the available budget predefined by the application owners. For example, in the case of the budget shortage, some levels of the performance degradations may be acceptable from owner's perspective. 

	\item \textbf{Root cause of the problem:} In traditional threshold based scaling, changes in the number of the VMs is the most common response to performance problems in the system. However, there are a wide variety of reasons from hardware faults to local software bugs in the application or security issues such as Distributed Denial of Service(DDoS) attacks that can create the signs of performance degradations. In cases that resource shortage is not the main reason for the problem, adding new instances to the system may temporally alleviate the problem, but it is not optimal as a long-run solution. Moreover, as the vertical scaling is becoming more prominent as a scaling option, having the knowledge of the underlying reason has become more interesting for a more cost or resource effective solutions. For example, in the case of a local memory shortage in one VM, a VM-level increase of the available memory may be more effective than adding new VMs. A more detailed explanation of the pro and cons of these types of decisions are presented in section \ref{actions}.
	
	\item \textbf{SLA agreements}: SLA agreements are contracts between users and CSPs which identify the expected QoS received by customers. These expectations are usually based on the outputs of the system perceivable by customers such as the availability of the service or the delays in the response. Having specific requirements for the output of the system may limit available choices of the RMM. For example, CSPs may consider over-provisioning as a better option than dynamic scaling to manage high loads in the system when having a stable response time is highly important for the customers.
	
	\end{itemize}  

\subsection{Data Aware Resource Management}
Motivated by the aforementioned challenges and requirements, researchers are leveraging various tools and concepts to offer more mature solutions for cloud resource management. An area which has been vastly investigated is data analytic techniques which are bringing new opportunities and challenges in the area of distributed performance management. In order to apply these techniques, researchers are focusing on the obtainable knowledge from the data collected from performance indicators of the system and applications. It has been shown that these data are a valuable source of information on the health of the system and a starting point for detecting initial symptoms of the abnormal behaviors. Based on the selected performance data to be monitored, the approach for the abstraction and modeling of the system and the actions that are performed to mitigate the performance problems, different types of the resource management strategies are proposed. In order to better understand the building blocks of these solutions, we first briefly review four main components of data aware performance management framework in the following paragraphs.

\begin{figure}
	\centering
	\includegraphics[scale=0.55]{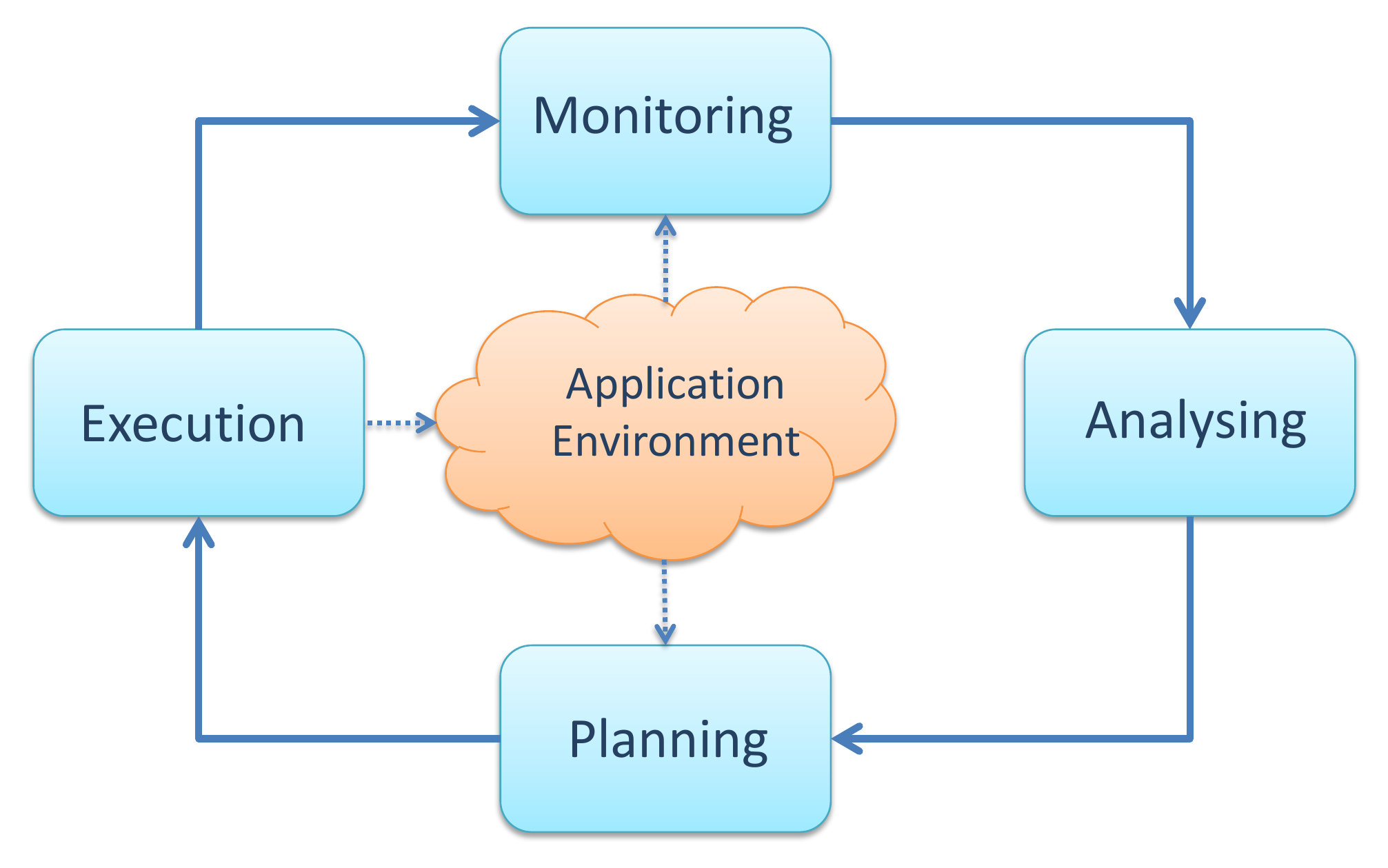}
	
	\caption{\label{fig:Gframe}
		General Phases of Data Aware Performance Manager in Cloud}
\end{figure}

The main parts of the automated resource management in the cloud can be explained based on a classic control loop known as MAPE (Monitor, Analyse, Plan, Execute) loop \cite{computing:2006:architectural} which is shown in Figure \ref{fig:Gframe}. The performance of the system is continuously monitored and a range of attributes from resources and applications are collected. The collected data are cleaned, modelled and analyzed to identify any symptom of changes in the normal behaviour of the system. Finally, based on the output of the analysis phase, a proper action is selected and the target components are informed to start the execution of the action. We briefly explain each phase and list the related categories of the taxonomy to each part in the following paragraphs. 

\textbf{\emph{Monitoring}}: The performance of the system can be tracked by collecting the values of the attributes from the components of the system. These attributes include all the workload metrics, system traces, network features or performance indicators of the system such as CPU and memory utilization, the number of incoming requests, the number of threads or response time of the application. The \textbf{data level} part of the taxonomy indicates different levels of the data collected during monitoring phase. 

There are a variety of tools to help monitoring and collecting data from system components including \textit{Top} and \textit{Iostat} packages or \textit{Ganglia} framework \cite{Ganglia:2018}. One can select a proper tool based on the factors such as the granularity of data to be collected, the level of access to the system components, scalability and characteristics of the system.  

One point worth mentioning is how to select a proper monitoring interval time. The interval can be selected as small as 1 second or a large value as 1 hour. Smaller intervals make it possible to capture the fast changing patterns or fluctuations with higher accuracy. However, the amount of the storage required for keeping all the recorded data and the overhead of processing and cleaning of the data significantly increases. Selecting larger intervals reduces the overhead and required storage, but the possibility of missing or delayed detection of changes in the pattern of the performance data increases which can cause delayed triggering of corrective actions and more SLA violations. Therefore, one should select a proper interval considering the trade-off between accuracy and computation complexity, the reliability of the environment and type of the application.

\textbf{\emph{Analyzing}}: In general, two main blocks of the data analyzer module are data preparation and performance modeling/analyzing. Therefore, all the steps required for cleaning and filtering, dimensionality reduction, building the models, analyzing new observations and deciding on the model updates when the state of the system changes are parts of this phase. A wide range of techniques and algorithms can be used to learn a model based on the historical behaviour of the system. The \textbf{Data Learning Approach} and \textbf{Performance Analysis Approach} parts of the taxonomy present different categorization of existing methods for this phase.

\textbf{\emph{Planning}}: 
The inputs for the planning module are the information about current or future state of the system from the analyzer, current configuration of the resources from application environment and the objectives and constraints from customers or resource providers. Depending on the obtained knowledge, the module can select from a range of possible actions such as adding/removing VMs, changing the configuration/placement of multiple VMs or inbound traffic balancing. Decisions can be formulated based on the past experiments and knowledge about possible causes of changes in the system. Therefore, the process can be implemented as a simple sequence of If-else rules or at a larger scale, as a database that can map a combination of influential parameters to their corresponding mitigation action. The subcategory presented in Figure \ref{fig:actions} focuses on the this phase. A detailed explanation of possible actions can be found in section \ref{actions}.

\textbf{\emph{Execution}}: This is where the final execution of planned actions in the system is performed. The module utilizes existing libraries and APIs to communicate with application components or deployed VMs to add new resources, remove the idle ones, change the configurations of existing VMs or updating load balancer configuration files. This phase more concerns the development strategies and techniques which are out of the scope of the current research. 

In the following sections, we present different aspects of a data aware resource management solution based on the categories shown in the Figure \ref{fig:tax} and subcategories presented in \ref{fig:sources} and \ref{fig:actions}. Based on the categories and identified approaches, we map each work to corresponding features in Table \ref{tab:title} to give the readers a quick view of the main contributions of each work.

\section{Performance Management in Cloud}\label{BDPM}
  Monitoring tools collect a valuable source of the data to be analyzed and provide a timely update on the performance state of the application and resources. Data learning approaches offer the researchers all the necessary concepts and tools to sift through the collected data and predict the future behaviour or find interesting patterns of unexpected behaviors or anomalies with their possible causes. In this section, two main approaches for analyzing the performance of the system are presented. 
 
\subsection{Workload-driven Performance Management}
Performance of the system can be modelled and predicted based on the workload-related features such as the number of requests received or amount of processing required at each time interval. \citeauthor{di:2014:google} propose a method for long-term load prediction in Google data centres. They consider load in the system as the main factor affecting the performance of the system and ignore other sources of data. In order to have a better representation of the statistical properties of the load including trends and seasonality, different metrics based on the load measurement values are derived. The prediction is done by training a Bayes classifier and exploiting a time window approach which is a suitable way to smooth high fluctuations in the load. However, other types of anomalies which can be directly related to specific performance metrics can not be detected in this approach meaning that unexpected behaviour can occur in the system, possibly causing negative impacts on the user experience. Work presented by \citeauthor{cetinski:2015:ame} consider a single attribute, number of required processors at a certain time, to estimate the utilization of resources. They expand the training dataset by introducing new attributes based on similar patterns in historical data. The results show that these new attributes improve the prediction accuracy of Random Forest compared to K-Nearest Neighbor algorithm. However, their prediction does not include the concept of unexpected behaviors resulting from various anomaly sources. VScaler proposed by \citeauthor{Yazdanov:2013:VAV} leverages a combination of workload prediction and reinforcement learning (RL) to automatically scale VM resources considering the user-provided SLAs. RL approach in this framework helps to automate the learning process, considering the uncertainty of environment in the form of changes in the workload model of the application. Another work by \citeauthor{Yang:2013:WPA} presents a cost aware resource auto-scaling mechanism which considers both costs of adding new VMs as well as business software license during scaling up procedure. A combination of linear regression based workload prediction, integer programming based pre-scaling and threshold based real-time scaling is introduced for capacity planning and resource management. The real-time scaling can be considered a reactive step to compensate prediction errors, but the simulation based validation of this approach ignores many complexities and time requirements of mentioned methods and these assumptions must be carefully verified.

In the workload explained performance management approach, the changes in the pattern of the workload are the primary influential factor that can affect the performance and hence the resource decision making of the system. The definition of the workload is dependent on the application and can be demonstrated as the number of requests sent to an interactive application such as web based systems, number of tasks/jobs running in the system and etc. One can also consider the resource demands of the jobs to be processed at each time as a representation of the existing load of the system. However, this assumption should be verified so that the resource consumption is a sole function of the load of the target application and all the dynamic factors such as the effect of background applications and sharing of resources are considered in the process.

\subsection{Anomaly Aware Performance Management}
A different approach to address the problem of performance management targets the abnormality in the system behaviour as a starting sign for possible problems to be addressed by resource management module. In the context of big data enhanced solutions, these performance problems can be seen as outliers and anomalies in the data which can be identified by utilizing a variety of methods such as statistical or machine learning algorithms. Therefore, we first define the term of anomaly in a general context; Then, the problem of identifying anomalies in the context of cloud is explained and existing works that follow this approach are discussed in more details.
\subsubsection{What are Anomalies?}
Anomalies are the patterns in the data that do not conform to the usual behaviour of observed data. The concept of anomalies and anomaly detection area are very general and presented under different names including outliers and novelty detection, finding surprising patterns in data, fault or abnormal behaviour detection in the systems \citep{Chandola:2009:ADS}. These areas have been investigated over a long period of time as part of the medical and clinical data clustering, image processing and surveillance cameras, financial fraud detection, and several other applications. 

\subsubsection{Performance Anomaly Identification in Cloud Environment} 
In general definition of anomaly detection, the goal is to model the normal behaviour of the system, so any unexpected change in the patterns can be seen as an anomaly. However, considering the user centric approach in resource management decisions in the cloud, the application owners are more interested in the events that can affect the performance of the system and degrade the quality of service experienced by the user. We refer to all of these events as performance anomalies. Considering that performance degradations can cause resource wastage, loss of reputation and cost penalties for cloud service and resource providers, many researchers have investigated the relation between measurements from the system and application-dependent performance indicators to have a better understanding of different causes of performance problems. We identify three levels of knowledge obtained from the process of performance anomaly analysis as shown in Figure \ref{fig:pslit} which are detailed in the following paragraphs:

\begin{figure}
	\centering
	\includegraphics[scale=0.55]{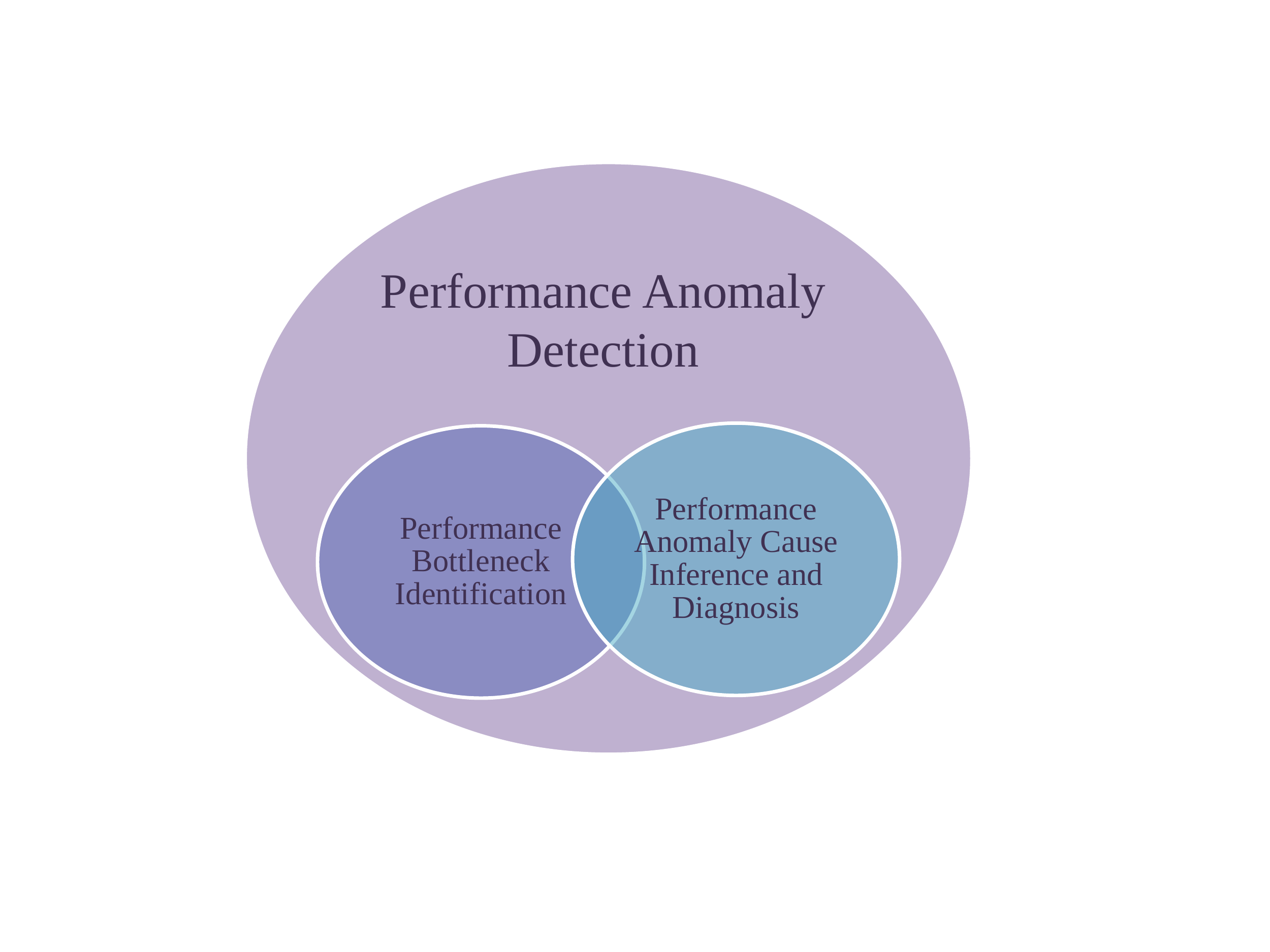}
	
	\caption{\label{fig:pslit}
		Different levels of knowledge from performance anomaly analysis}
\end{figure}

\textbf{\emph{Performance Anomaly Detection}}
The goal of a data analyzer module at the performance anomaly detection level is to find any abnormal pattern in the behaviour of the system that can be a symptom of the performance problems. Therefore, the input for these frameworks is usually the system and application performance indicators while the output is a performance alert when an anomaly is detected in the system. 
Having this goal and considering the fact that a correlation of different metrics can be related to various types of anomalies, \citeauthor{Guan:2013:AAI} present an automatic anomaly identification technique for adaptively detecting performance anomalies such as disk and memory related failures. Proposed method investigates the idea that a subset of principal components of metrics can be highly correlated to specific failures in the system. A combination of Neural Network method and Adaptive Kalman Filter is utilized during a procedure of learning from historical data, updating prediction model based on the current prediction errors and adapting to the newly detected anomalies to improve the detection performance. \citeauthor{Cunha:2012:SPA} focus on two general categories of anomaly sources, workload related and performance related data in streaming servers. They justify this separation as a requirement to select the best repair action in the response to degradations caused by targeted faults. A feature selection procedure based on Naive Bayes is employed and the most relevant features are reported.
\citeauthor{Ashfaq:2014:ITFS} target the problem of anomaly detection from a new perspective, highlighting the scalability problem of data analytic solutions for resource management issues in the cloud environment. They propose a general framework for anomaly detection based on the splitting feature space into multiple disjoint subspaces and applying anomaly detection methods on each subspace separately. The idea behind using feature space slicing is to decrease the likelihood that a high number of normal instances can average out the effect of a few dispersed numbers of malicious instances during anomaly identification. Since this approach requires higher computation resources, as it needs to run multiple simultaneous instances of the algorithm on different subspace, it is more suited for high performance computing platforms.

BARCA (Behaviour Identification Architecture), proposed by \citeauthor{Cid:2015:OBI}, is a framework for online identification of anomalies in distributed applications. It divides the anomaly detection process into two steps. First, a one class classifier is employed to distinguish normal behaviour from unexpected ones. Then, a multi-class classifier is used to separate different types of abnormal behaviours. The framework generates time series of different collected performance data and extracts new features such as skewness and mean of data which better represent the characteristics of the time series and also help to reduce the dimensionality of feature space.

The above-mentioned works target the reactive anomaly detection problem in the cloud environment. In order to be able to move the system back from abnormal to the normal state as soon as possible with minimum negative impact, we need to know about the probability of changing system features to abnormal values in the future. In the proactive approach, systems are able to exhibit goal-directed behaviour by anticipating possible future abnormalities and taking initiatives \citep{Huebscher:2008:SAC}. \citeauthor{gu:2009:online} investigated proactive anomaly detection in data stream processing systems. Their proposed solution includes a phase of predicting resource utilization and then applying an anomaly identification algorithm on predicted data. Considering time sensitiveness of stream data, proposed procedure is online and the classifier will be updated periodically based on the new data. To address the prediction problem, they apply Markov chain to capture changing pattern of different metrics to predict future resource utilization. The Markov chains are based on the idea that future state only depends on the current state and not the past values. This assumption can be problematic, especially for the data with recurrent patterns and events. \citeauthor{Tan:2012:PPP} address this problem by integrating a 2-dependent Markov model as the predictor with Tree-Augmented Naive (TAN) Bayesian networks for anomaly detection. Another study by \citeauthor{Dean:2012:UUB} investigates unsupervised behaviour learning problem for proactive anomaly detection. The proposed framework uses Self-Organizing Maps (SOM) to map a high dimensional input space (performance metrics) to a lower dimensional map without losing the structural information of original instances.

\textbf{\emph{Performance Bottleneck Identification}}
Performance bottleneck identification goes one level deeper in the process of finding anomaly events in the data, trying to find possible bottleneck metrics that are closely related to the observed performance degradations as well. This approach is closely related to the problem of the resource management as it targets finding possible system resources that need to undergo a reconfiguration so the provided resources meet the requirements of the application. \citeauthor{Tan:2012:PPP} leverage TAN to distinguish normal state from abnormal ones as well as reporting the most related metrics to each type of the anomaly. Canonical correlation analysis and Support Vector Machine (SVM) based feature selection are used by FD4C, a framework presented by \citeauthor{Wang:2016:AFD} to diagnose faults in the web applications. They utilize a recursive approach based on the feature elimination to rank the most important metrics for each type of the anomaly. \citeauthor{Xiong:2013:VAM} have a different approach for detecting the performance bottlenecks. They try to find the most relevant metrics to the performance of the application and follow the changes in these metrics as a sign of the performance problems. However, they show that the predicted metrics are also good indicators of the source of analyzed performance problems, pointing to the source host and type of the bottleneck resource. UBL presented by \citeauthor{Dean:2012:UUB} uses the topological properties of SOM to compare the anomaly and normal states and identify the metrics that differ most between these states as faulty metrics.

\textbf{\emph{Performance Anomaly Cause Inference and Diagnosis}}
The aforementioned anomaly detection approaches mostly focus on detection of the abnormal symptoms and a coarse-grained identification of the possible \textit{resource level metrics} that contribute to the performance degradations. However, none of them digs deep into the data obtained from the application to find the underlying reasons for the observed problems. Indeed, identified bottleneck metrics can be the indicators of having an application or VM level fault or inconsistency in the system. For example, high incoming load to the application or a faulty loop in the software code which saturates the CPU of the VM or the problem of VM/application contentions which may cause degradations in memory utilization. 

We can identify different directions in the fine-grained analysis of the source of the faults. First, the works that aim to localize the source of the fault to one \textit{component} such as nodes, VMs or application components. For example, the works done in \cite{Nguyen:2013:fchain, Nguyen:2011:PPR} address the fault localization problem in distributed applications. The proposed frameworks combine the knowledge of inter-component dependencies with change point selection methods, taking into account that abnormal changes usually start from the source and propagates to other non-faulty parts based on the components interactions.

Another direction is to distinguish among different \textit{types} of the faults. \citeauthor{Dean:2014:PTR} propose PerfCompass which analyzes the generated system calls in the system to distinguish between internal and external faults. They focus on software related bugs such as endless loops as the target internal faults. \citeauthor{Cid:2015:OBI} apply a set of the SVM based binary classifiers to distinguish among livelock, deadlock and starvation faults.

 To achieve a more fine-grained identification of the cause, \citeauthor{Dean:2014:PPO} propose PerfScope to analyze the anomalies occurring due to the software bugs of the application. The framework studies the patterns in the \textit{system calls} and tries to find anomalous interactions between user and kernel. Triage, proposed by \citeauthor{Tucek:2007:TDP}, is another online software failure diagnosis which identifies the conditions as well as the \textit{code} and variables involved in the failure state. TaskInsight presented by \citeauthor{Zhang:2016:TI} focus on the thread and process level performance information which helps to localize the problem to the target anomalous task.

\begin{figure}
	\centering
	\includegraphics[scale=0.95]{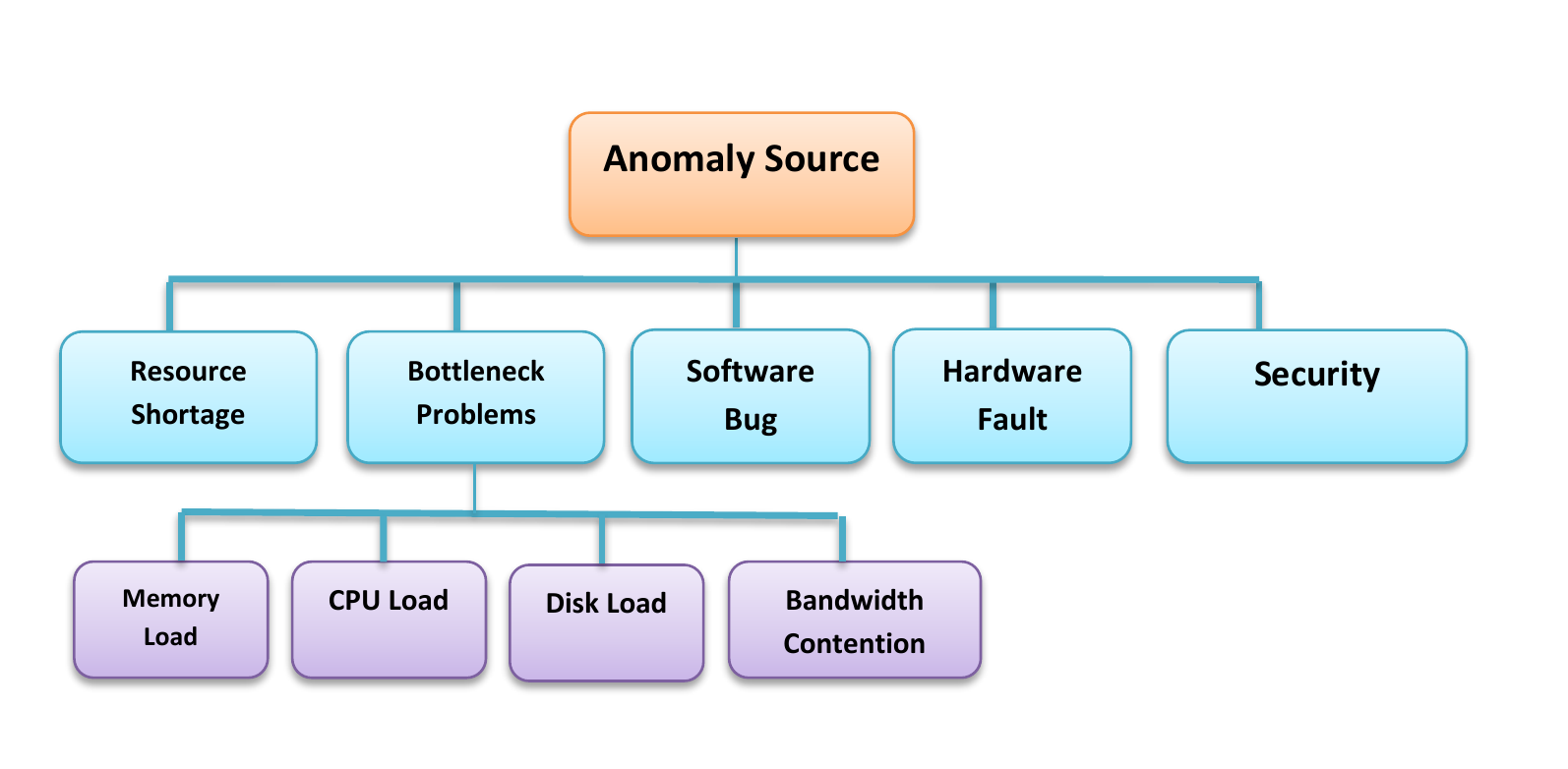}
	
	\caption{\label{fig:sources}
		Source of Performance Problems}
\end{figure}

\subsubsection{Cloud Performance Anomaly Root Causes}
Cloud application owners typically start to allocate resources based on the recommended application requirements and then change resource configurations by continuously monitoring performance indicators to find performance violations. The root cause of these performance problems can vary widely as shown in Figure \ref{fig:sources}.

 Hardware faults include the problems that are originated from the corrupted or performance degraded hardwares that host target applications \citep{Do:2013:LUI, Vishwanath:2010:CCC}. Software related problems can be caused by a buggy code in the application or misconfiguration that causes inconsistency in the functionality of software or interactions among components. This type of problem also can be caused by network related bugs and misconfiguration that happen at the application level, including reported bugs in skype, MySQL or IPv6 compatibility issues \citep{Zhuang:2014:NND}. 
 
 Security problems including attacks are another source for the unexpected behaviours caused by the unusual pattern of requests such as successful port-scans, attacks on the application server, etc. A wide variety of literature target this area, proposing various types of intrusion detection systems based on the concepts of statistical feature analyzing, classification and clustering \cite{Ashfaq:2014:ITFS, Lin:2015:CANN, Hatef:2018:HIDCC}. In order to identify these types of the problems, one needs to collect network layer datasets including packet header information or the frequency of sender IP addresses to detect unusual patterns in the requests \cite{Ashfaq:2014:ITFS, Shamsolmoali:2014:SFS}. Other sources of data to help recognize the access patterns to the application are server log files which record the history of authentication and user access requests over time. 
 
 Resource shortage issues are another reason for the performance problems when the lack of enough resources to satisfy existing requests causes service interruptions and degradations in the performance. The limitation can be due to the budget constraints or business policies that do not allow adding extra resources to the system or reduces the amount of the existing resources. Unexpected termination of instances performed by resource providers is one case of these problems that can expose the system to a state that the performance and throughput of the system will be degraded \cite{Qu:2016:RCA}. 
 
 Bottleneck problems are another reason for the performance problems which are caused by insufficient resources in one or more components of the application. The problem can be due to the specific requirements of offered services such as working with a CPU intensive software and the lack of consistency among the application demands with provided resources. Another example is the effect of the background processes which can temporarily saturate the resources of the machine, ignoring the requirements of other installed applications. It is worth noting that if the components are dependent and have interactions, the bottleneck problem in one part of the system can quickly affect other dependent components and as a result, the performance problem will propagate in the system causing application performance degradations \cite{gu:2009:online}.

\section{Data Level} \label{Data_Level}
The monitoring module can collect data from a variety of system resources and application components. In general, we have identified four primary sources of data that give information on the system from different perspectives to be processed by the analyzer.

\emph{\textbf{System-Level}}: System level metrics refer to the collection of attributes that can describe or predict the behaviour of running environment including VM or physical machine. One category of these attributes is resource level metrics which act as the performance indicators of the running system at different levels of granularity from VMs to specific process and threads. For example, one can present the number of assigned CPU cores or the percentage of used CPU, Memory or Disk I/O at different time intervals as indicators of the functionality of the system during runtime of the applications. Another source of data which can be categorized as part of the system metrics are generated system calls which show the pattern of interactions with operating system services. It is shown that these patterns can be affected by different types of internal and external faults which help to detect and localize the source of the faults \cite{Dean:2014:PTR, Dean:2014:PPO}.
 
 Many cloud resource providers offer monitoring services to collect data from physical and virtualized machines. Amazon CloudWatch is an example of these services. There is a range of system monitoring and debugging tools such as htop, Iostat and strace, each offering a level of information about utilization of resources or pattern of interactions among processes in the system. While these tools give valuable information about the functionality of one machine, their usage for a cluster of machines needs more scripting and data management. For a more flexible monitoring of the distributed systems, advanced frameworks such as Ganglia are introduced \cite{Ganglia:2018}. Ganglia is a distributed framework based on the hierarchical design which uses technologies such as XML and RRDtools for monitoring different components of the system. It is accompanied with a dashboard to view live statistics of the monitored system while the recorded data is also available for deeper analysis.

One point worth mentioning here is that the above-mentioned tools require one to directly access the monitored VMs, or some components of the monitoring modules should be installed on the machines beforehand. Therefore, these approaches are suitable for cloud application owners who have access to the VMs. There are also works that follow the blackbox rules, trying to avoid or decrease the dependency on the application or guest VMs by utilizing hypervisor capabilities to collect data from outside of the VMs \cite{Tan:2012:PPP, Islam:2012:EPM, cetinski:2015:ame}.

\emph{\textbf{Application-Level}}: Application level metrics are collected from application components deployed in the system. Since these metrics are directly related to the runtime state of the application, they can be very informative and good indicators of the health of the application or environment. For example, in a web based application, response time which is the delay from initiating the request until the user receives the results is considered as an indicator of the quality of the offered web service. Longer response times can be warning signs for a high number of requests or some problem inside the systems \citep{Yang:2013:WPA, Tan:2012:PPP}. There are also more fine-grained works that study the software code and debug the flow of data, compare the effect of various input variables or environment configurations \cite{Tucek:2007:TDP}. These works are more related to the field of software debugging or runtime diagnosis. However, we have included them in our survey as they can help to distinguish internal application faults from external ones which consequently affects the type of the corrective actions from service provider side.

\emph{\textbf{Network-Level}}: There is a body of works which study the obtainable knowledge from analyzing network level data such as packet headers or the frequency of received packets from specific users which makes them suitable for identifying network related issues and particularly security threats such as DDoS attacks. \citep{Ashfaq:2014:ITFS, Shamsolmoali:2014:SFS}.
The source of the problem in these cases is usually associated with the external factors and therefore these frameworks are complementary to the ones that directly target the internal state of the system and resources. 

\emph{\textbf{Structural-Level}}: While a per component monitoring gives us valuable information on the functionality of individual parts of the system, in a distributed system, these components are in continuous interactions. In other words, the functionality of one part may be dependent on the correct execution of another part. Therefore, the fault in one component can be quickly spread based on the application architecture and the path of flow of the data/commands among components. Having the information on the execution order of application components along with the timestamped data of performance metrics can give new insights into localizing true sources of the faults which are propagated from different layers of the application \citep{Nguyen:2013:fchain, Nguyen:2011:PPR}.

\section{Architecture}\label{Arch}
In an environment with geographically distributed resources, the decisions on the placing and interaction among components can highly affect the performance of the system and corresponding corrective actions. Various factors such as the amount of available storage, resource demands or the speed of information dissemination can contribute to these decisions. In the following, we briefly discuss two main approaches for the placement of different components of a resource management framework in the environment.
 
\subsection{Centralized}
Traditional frameworks to analyze the health state of the system, typically follow a centralized approach. In a centralized structure, all data from local components including performance indicators and resource configurations are sent to a master node. The master node is responsible to maintain a continuously updated model of the whole system and trigger alarms when a performance problem is happening. All the tasks about workload prediction, resource utilization estimation or performance problem analysis is done at this module \citep{Islam:2012:EPM, Cohen:2004:CID}. 

An advantage of a master node approach is to have all system related performance information in one place; therefore, one can analyze the interactions and relation among the metrics at different layers and tracks the fault propagation among connected components. However, as the scale of the system increases, there will be more components and resources to be monitored usually in geographically distributed regions. The process generates a huge volume of collected data to be transferred and analyzed in one place. This makes the system modeling and abstraction as well as triggering a proper resource management action to be traversed across all the involved resources, super complex, computationally intensive and time consuming.

\subsection{Distributed}

In a dynamic and scalable system, administrators are more inclined to deploy computing modules as decentralized components \citep{Tan:2012:PPP, gu:2009:online, Nguyen:2013:fchain}. Therefore, each VM/physical machine in the system will have a local analyzer dedicated to processing locally collected monitored data to model the behaviour of the system. This approach can be easily deployed and has higher scalability and manageable computation time. Moreover, failures in one node do not affect the functionality of remaining processing modules.
However, the lack of a central manager makes it hard to include the interaction and dependency among components during analysis. Each module has an abstract performance model of its local environment, completely ignoring the effect of any external factors such as the dependency among different layers of multi-tier applications. As a solution for this problem, one can consider hierarchal frameworks, where each monitored component has a local analyzer module as well as the ability to send a summary of the local state of the component to a central node. Central node combines the knowledge from local states and the dependency among components to create a general model of behaviour for the whole system. This approach combines the scalability characteristic of the distributed methods with system wide knowledge of the centralized master node which helps the system to respond to local problems quickly while having enough knowledge to diagnose and plan for the global problems.

\section{Data Learning Approaches}\label{Data_Analysis}

The core part of the data aware resource management is the data analysis module which obtains the knowledge of the current or future state of the system to help decide the best action to keep the system compliant with the SLA requirements and business goals. There are different approaches to help learning and analyzing the health of the system from collected measurements. We categorize and summarize the characteristics of the identified approaches in the following subsections.
\subsection{Signature Based Analysis }
A state in the system can be characterized by the values of the attributes of its components at different levels of granularity. Considering that various types of the faults or performance problems leave distinctive signs on the attributes, one can capture a snapshot of all values during abnormal or normal behaviour and represent it as the signature or fingerprint of this state. The works presented by \citeauthor{MI:AAPC:2008} and \citeauthor{Cherkasova:2009:AAD}, distinguish performance problems caused by anomalies from the ones which are the results of an application update or the changes in the resource consumption models. They create a profile of the application performance based on the concept of transaction processing times and related resource utilizations. The profiles are used for the comparison between old and new application performance to detect the changes. While these works leverage the application related measurements to create profiles, \citeauthor{BRUNNERT:CPECPU:2017} utilize resource profiles to detect the performance changes in the enterprise applications (EA). The resource profiles are defined as the amount of required resources including CPU and memory for each transaction of an EA version. The independence of the resource profiles from hardware and workloads makes them a more flexible solution for areas such as capacity planning or energy estimation. There is also another approach which utilizes the concept of the signatures at the diagnosis levels. For example, \citeauthor{Sharma:2013:CPD} propose a fault management framework which first detects an anomalous behaviour by statistical analyzers. Then, detected deviations are matched with the pr-defined signatures of the faults to identify the cause of the problem. 

While the signature based approaches usually show low false positive rates, an important challenge is the creation of the baseline profiles which captures the target states of the system. Regarding anomaly detection problems, creating signatures need the domain knowledge of the problem and there is always a high chance of missing unknown anomalies which increases the false negative rates in the results.

\subsection{Threshold based Approach}
Threshold based approach is a simple yet popular way among service providers for defining a set of rules to manage the resources in the system \cite{amazon:2018}. The idea behind this approach is that the anomalies can cause unusual increase or decrease in the utilization of the resources, affecting the values of the attributes for the system or application. One can define the scaling up/down rules by identifying a threshold for the acceptable utilization in the system. If the target resources exceed the threshold, a scaling action should be triggered. Therefore, two main parts of each threshold based rule are the condition and the action. Regarding the condition part of the rule, an attribute of the system which can be a resource level metric or application level performance indicator is selected. Then, proper lower/upper thresholds are identified. Whenever a threshold value is exceeded, the conditions are met and the action is started. The second part of the rule is defining appropriate actions such as deciding on the number of VMs to be added or the VMs which can be shut downed in the system. This type of the action corresponds to horizontal scaling policy which is offered by most of the cloud resource providers. \citeauthor{Gmach:2009:RPM} investigate the reactive threshold based approach to detect over-utilized or underutilized servers. \citeauthor{Yang:2013:WPA} extend this approach with a linear regression based prediction phase and apply one of the vertical or horizontal policies when a violation of the threshold is met. There are also works which implement the threshold based policies for the baseline comparison with their proposed frameworks \citep{Han:2012:LRS, Hong:2015:DDA}. For example, \citeauthor{Hong:2015:DDA} compare Markov based anomaly detection scheme with a simple threshold based monitoring which triggers anomaly alerts when the resource utilization thresholds are violated.
\subsection{Control Theory}
Control theory helps to automate the resource scaling decisions by creating a systematic way of adapting to the dynamics of the system. The controller should trigger proper corrective actions by adjusting the values of input variables of the system to maintain the output or controlled variables close to a baseline. Regarding \textit{Open-loop controllers}, the corrective action is selected solely based on the inputs, while in the \textit{ closed-loop} also known as \textit{feedback controllers}, the changes in the controlled variable is received as the feedback to be considered by the controller for the next action. The feedback controllers are the most common approach in the target area for this paper. \citeauthor{Grimaldi:2015:FCA} propose a Proportional-Integral-Derivative (PID) based controller by managing the number of VMs in the system, aiming at keeping the service quality in accordance with the agreement levels. Integration of the Kalman filter and feedback controllers are investigated by \citeauthor{Kalyvianaki:2009:SSC} to manage the allocation of the resources based on CPU utilization of the VMS. \citeauthor{Al-Shishtawy:2013:EEM} combine both feedback and feedforward controllers, harnessing the power of both approaches for multi-tier applications. Feedforward part is acting as a predictive controller to proactively avoid SLA violations caused by unexpected increases of the workload. In the case of the violations, feedback controller reacts to compensate the deviations in the performance.
\subsection{Statistical Approaches}
Statistical approaches usually assume that the key attributes of the system follow a known or inferable behavior. Therefore, observing and collecting the data on the system attributes over time provides a baseline which any deviation from that is identified as an anomaly. The definition of the baseline behaviour is usually based on some statistical characteristics of the data such as mean and standard deviations. For example, many works on the anomaly detection are based on the assumption that values of the target attributes follow a normal distribution. Multivariate Adaptive Statistical Filtering (MASF) is a common method in this group which tries to find multiple sets of the control limits based on the statistical analysis of the previous measurements of the features during normal system operations \citep{BuzenS:1995:MASF}. The observations outside of the control limits are considered as possible anomalies in the system. \citeauthor{Wang:2011:SAD} generalize this concept to more flexible thresholds, being more adaptable to dynamics of the workloads in data centers. 

While these solutions are simple and lightweight, the highly dynamic nature of the cloud requires more flexible solutions which can capture the relation among features. A body of works \cite{Ibidunmoye:2017:BDS, Peiris:2014:PAD, Matsuki:2016:RCA} try to show some type of the correlation among resource metrics and QoS indicators, using this information for better understanding of the nature of the anomalies and further performance analysis such as cause identification. Another approach which addresses the problem of proactive resource scaling leverages regressions based methods to find the relation among metrics and performance indicators \cite{Islam:2012:EPM, Xiong:2013:VAM, Yang:2013:WPA, Wajahat:2016:MLA}. The prediction of the future workloads or resource utilization provides the planning modules information on the possible changes in the system that require a reconfiguration of the resources. For example, MLscale is an auto-scaler which uses the regression method to predict the values of the metrics and consequently the performance indicators of the system through a hypothetical scaling and decides on the best action based on the results \citep{Wajahat:2016:MLA}. 
 
Another area within the domain of distributed resource management, where statistical techniques have been commonly used, is analyzing network level state of the system which helps to distinguish between normal traffic and network related security issues such as attacks. \citeauthor{Gu:2005:DAN} employ relative entropy to compare the new traffic data with the baseline distribution and identify anomalous traffic. \citeauthor{Cao:2014:EDS} target DoS attacks launched by malicious tenants of the VMs in cloud data centers. Entropy is calculated based on the resource and network utilization information. They show that the entropy of VM's status drops when the attack starts. \citeauthor{Ashfaq:2014:ITFS} divide the feature space of the problem based on the information content concept, putting statistically similar instances in the same subspaces. The idea behind this approach is to avoid the effect of averaging out of anomaly points and also localizing noise artifacts in separate subspaces.

\subsection{Machine Learning}

Machine learning concepts include the techniques that enable a system to learn from the experience over time without being explicitly programmed. The massive amount of the collectible data from the system is a valuable source of the information to be utilized by these techniques to learn from the environment. Each technique tries to structure data in a different way to generate an abstract model relating the input to output variables. Generally, we can divide these techniques into two main categories Supervised and Unsupervised. In the following, we explain each category in more details.
\begin{itemize}
	\item \textit{Supervised Learning}
	Supervised algorithms require the dataset to be labelled, meaning that the desired output should also be clear during the training phase. This approach is more suitable for problems that the goal is to find a mapping between the input and output variables, so having the new input observation one can find the possible output. A common case of utilizing this technique is the classification of a set of records when each input should be assigned to one of the predefined classes. In the area of anomaly detection techniques, the classifiers can be used to categorize different types of anomalies or more generally distinguish between normal and anomaly state of the system. Following this approach,\citeauthor{gu:2009:online} try to detect the type of the future anomalies by using naive Bayesian classifier. A set of binary classifiers are trained to distinguish among different types of bottleneck problems. The authors show that the proposed classifiers can achieve high accuracy, detecting the anomaly symptoms caused by some of the common bottleneck issues at the application and resource level. Similarly, \citeauthor{Tan:2012:PPP} exploit TAN to predict the anomaly state of the system. 
	\citeauthor{Cunha:2012:SPA} apply two classification algorithms, J48 Trees and Naive Bayes, on the historical data through ten-fold cross-validation. The goal is to differentiate between workload related anomalies that are caused by higher request rates and other types of performance anomalies including memory leak in streamed video data.
	
	Neural Networks are commonly applied for the prediction of the future utilization, performance indicators or workload metrics to model the performance of the application \citep{Islam:2012:EPM, Ajila:2016:MLCP, Wajahat:2016:MLA}. Rather than a direct identification of anomaly solely based on the context of the data and patterns, these works usually focus on finding the symptoms of the performance degradations in the application. In contrast, \citeauthor{Guan:2013:AAI} identify anomaly events by combining neural network and kalman filters to adaptively calculate the principal components of data. The idea behind their approach is that a subset of principal components is more related to specific types of the failure in the system.

	 The aforementioned works are trained to find a relation between input and output variables to predict output values for test instances. The outputs can be related performance metrics such as response time or a category of anomaly states assigned to the input instance. A major limitation of these works is the requirement to have a labelled dataset for the training. The process of labelling a dataset is time consuming and needs a good knowledge of the domain problem. Especially, in a dynamic environment, there is always a chance that the underlying mapping of the variables changes which require continuous reconfiguration and regeneration of the models for the new states of the system.

	\item \textit{Unsupervised Learning}
	In contrast to the supervised learning, unsupervised approach sifts through data trying to find hidden structures and patterns. Therefore, it does not need any prior information about the labels of training data. The objective is to cluster input data based on their features without any assumption about their distribution \cite{Zhang:2016:TI}. The unsupervised learning is particularly suited for the cloud environment, where the system administrators may not have access to the detailed VM utilization states or be unaware of the internal performance of the application. Following this approach, \citeauthor{Dean:2012:UUB} propose Unsupervised behaviour Learning (UBL) and investigate unsupervised learning problem for proactive anomaly detection. UBL is a framework which applies Self Organizing Map (SOM) to identify the anomalous states in cloud systems. SOM is an unsupervised type of the artificial neural networks that projects data instances from a high dimensional space to a lower space (usually two) while keeping the topological structure of data. Comparing the neurons of the generated map, they distinguish normal and anomaly states while a list of ranked metrics can also be inferred as a starting hint to find the cause of the problems. Hidden Markov Models (HMM) are used by \citeauthor{Hong:2015:DDA} to model system as a Markov process. In a Markov process, it is assumed that the state of the system at each time is only dependent on the previous state. Two hidden states normal and anomaly are determined and the probability matrices are initialized through an unsupervised training process.
	A two-phase clustering approach is proposed by \citeauthor{Li:2015:SCVM} which tries to find a VM placement solution to minimizes the number of physical machines as well reducing the performance degradations caused by the contention between co-located VMs. Hierarchical and K-means clusterings are applied to cluster VMs first based on the peak of the utilization and then the correlation among them.  
	\citeauthor{Ashfaq:2014:ITFS} apply clustering as a preprocessing phase to divide feature instances into distinctive categories based on their statistical attributes. Consequently, these clusters form the basic blocks of data to be analyzed separately by anomaly detection modules.
	
	Unsupervised learning helps the system to be able to detect both known and unknown anomalies. The process does not assume any prior knowledge about the statistical features or patterns in the data and tries to find the common characteristics observed among different sets of the instances. However, depending on the level of the details provided in data, the accuracy of unsupervised learning to recognize the exact categories of anomalies may be affected.

\end{itemize}

\subsection{Reinforcement Learning}
	Reinforcement learning (RL) focuses on the gradual learning through sequential interactions of the agents with the environment. The target goal of the agent is to maximize a reward function by selecting the best possible action based on the state of the system. The important feature of this approach is learning by experiences from the environment which helps to start the process without a prior knowledge of the system. In the area of cloud resource management, auto-scaler can act as an agent which interacts with system components including VMs and physical machines. The state of the system is represented by the system attributes and performance indicators, while the reward is shown by the degree of QoS (such as response time or throughput) achieved by the application. The set of actions include all possible corrective actions such as resource and application level reconfigurations to avoid the performance degradation. \citeauthor{Dutreilh:2010:DCR} compare threshold based with Q-learning approach for the problem of horizontal auto-scaling in the cloud. They investigate the functionality of each method in the presence of the main instability sources in the control systems, listing the observed potentials and weak points for each case. VScaler proposed by \citeauthor{Yazdanov:2013:VAV} is another framework for the fine-grained resource management in the cloud which utilities RL to decide on the times to scale up/down in the system. To speed up the process of learning and exploration of the controller, they introduce the parallel technique which enables multiple agents to collaborate in different parts of the system state space. \citeauthor{Duggan:2016:RLA} target the problem of VM live migration considering the available bandwidth and network congestion problem. They formulate the problem as an autonomous control system through utilizing RL and creating a multidimensional state/action space based on the VM utilization and available bandwidth. \citeauthor{Arabnejad:2017:CRL} extend a rule-based fuzzy controller with 2 different RL approaches, Q-learning and SARSA. The fuzzy concept helps to reduce the dimensionality of the state/action table which is an important issue affecting the complexity of the RL algorithms.
	
	As we can see, the gradual learning concept provided by RL fits the nature of the problems in cloud performance management very well by involving the dynamism and uncertainty factors during the learning procedure. However, one main challenge that RL based solutions suffer is the size of the possible states and actions for the system. Considering the continuous nature of time series measurements and the scale of the target machines to be handled in large scale distributed environments, the problem of high dimensionality is becoming more important. To overcome this limitation, different approaches such as fuzzification of the table or using a more abstract representation of data to limit the possible states or actions are proposed in the literature \cite{Arabnejad:2017:CRL, Liu:2017:AHFPM}.

\section{Action Trigger Timing: Proactive vs Reactive} \label{Decsion_Timing}
When a corrective action should be triggered is a challenging question as it is highly dependent on the nature of the application and SLAs. Traditional approaches to this problem are mostly reactive. In reactive methods, any decision about the changes in the number of VMs, the configuration of resources or VM replacements is a response to the abnormal behaviour of the system indicators that identify changes in the performance or quality of services. As the degradation already has occurred and considering the delays before corrective actions take effect, an amount of the SLA violations should be acceptable and taken into account in SLA contracts. Moreover, in an unreliable environment where there are various factors that can affect the stability of the system, fluctuations in the performance are a common observation which can increase the number of occurrences of SLA violations. Usually, these approaches follow a threshold based strategy where the scaling starts after the value of some metrics exceeds an accepted value  \citep{amazon:2018, Iqbal:2011:ARP}. For example, when the CPU utilization of server exceeds the threshold, new resources are added to the system.

 In contrast to reactive methods, proactive approaches attempt to find the warning signs before they can cause unacceptable levels of performance degradations, so they can start preventive actions such as migrating VMs or adding new resources \citep{Tan:2012:PPP, Yang:2013:WPA}. Proactive approaches focus on the future events and are mainly based on the prediction of the future values and states. If proactive analysis of data can give an early enough alert of a possible performance problem, it helps RMM to plan and quickly start a proper action before the system goes into an anomaly state. Accordingly, the system returns or continue the normal condition, reducing the number of violations. One point worth mentioning here is how to decide on a proper value for the period of prediction. Short-term predictions are more accurate in the case of the workload related metrics because the measurements close in time show higher correlations compared to observations for longer in time. Longer time predictions are more challenging and better fit the data with regular patterns or seasonality. \citeauthor{di:2014:google} investigate long-term prediction problem for workload data in cloud data centers. They propose a Bayes model based method to predict the average load in the system, based on the derived features that capture different aspects of the statistical characteristics of data. Another strategy to decide on prediction interval is considering the required time for the corrective action to be effective in the system. If the workload is predicted for an interval shorter than the action time, the value of proactive resource management diminishes. \citeauthor{Islam:2012:EPM} follow this approach and determine a prediction interval based on the time it takes to launch a new VM instance. Therefore, upon receiving an alert of a possible load problem, the system has enough time to start the scaling action.

\begin{figure}
	\centering
	\includegraphics[scale=0.60]{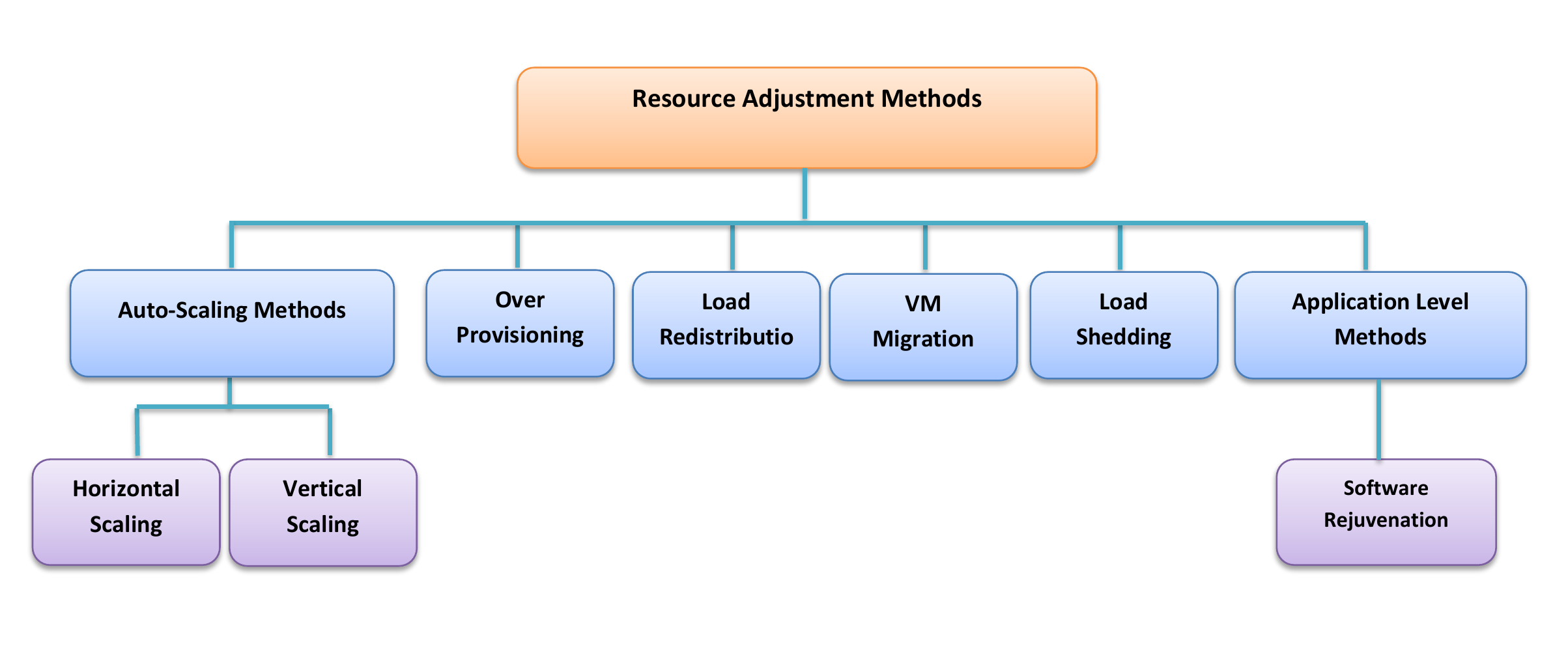}
	
	\caption{\label{fig:actions}
		Cloud Resource Adjustment Techniques for Performance Management}
\end{figure}

\section{Resource Adjustment Methods} \label{actions}
Upon receiving an alert of an ongoing or possible performance problem in the system, RMM should start an adjustment process so the active resources meet the storage and computing requirements of the workload of the application. There are different solutions to alleviate the performance problems in the system including changing resource configurations, adding new computing units or replacement of the VMs. In this section, we explain these techniques more and especially focus on coarse-grained methods at resource and VM level as shown in Figure \ref{fig:actions}

\subsection{Application Level Methods}
This category of actions are directly applied to the application and its environment, and therefore they are not specific to cloud systems. This approach targets the degradations in the performance of the application or operating system caused by the internal problems such as data corruption, numerical error accumulation or exhaustion of operating system resources \citep{Vaidyanathan:2005:CMS}. Software rejuvenation is a possible corrective action in these types of the problems which tries to identify the problematic application or system component, clean its internal state and restarts the components. Indeed, simple application or VM restarts which are used by the system administrators as the first reaction to many cases of performance degradations are preliminary cases of applying this approach.
 
\subsection{Over-provisioning}
In order to efficiently utilize the capacity of cloud offered resources, it is vital to have a proper estimation of required resources that keep the performance and QoS of applications at an acceptable level. However, resource estimation is a complex problem especially for dynamic workloads with time-dependent fluctuations. A traditional solution to handle the uncertainty in resource demands of the applications is to provide enough resources to process the maximum expected workload in the system. This solution guarantees a stable application performance in the presence of workload fluctuations. However, the peaks in the workloads are usually transient events that do not last long and happen very rarely which means that most of the times resources are under-utilized and incur extra costs for CSPs. Considering dynamic nature of the applications such as web workloads, over-provisioning is not avoidable in the fault tolerance resource management solutions; However, researchers try to find a proper trade-off between the level of fault tolerance of the system and over-provisioned resources to have more control over the functionality of the system \citep{Qu:2016:RCA}.

\subsection{Auto-Scaling Methods}
Scaling is defined as increasing or decreasing the amount of resources (ex: CPU, RAM, Disk, etc) for meeting SLA and performance standards. The scaling of resources can be done by changing the number of machines in the system or at the VM level by changing the configuration of one VM or. In the following paragraphs, we explain each of these approaches in more details. 

\subsubsection{Horizontal Scaling}
 Horizontal scaling which is the primary block of every RMM framework helps to provide more resources by adding new VMs in the system \citep{Iqbal:2011:ARP, Yang:2013:WPA, Yang:2014:CAA}. The unit of changes in horizontal scaling is one VM. However, the newly added VMs can have a customized resource configuration (CPU, RAM, I/O) or a pre-configured VM can be launched by selecting one of the instance types offered by the provider. For example, Amazon offers a range of VM types with different levels of configurations from small to very large instances \cite{amazon:2018}. As a complementary option, Google also offers users the chance to define their requirements with more details by launching customized instances. 

The time it takes for a new VM to be launched, also known as VM start-up time, is highly important, especially for proactive RMMs. If start-up times are longer than the predicted point for a possible breach of the SLA, the effect of the scaling process will be same as the reactive strategies while the overhead of data analysis module is yet added. Considering this, \citeauthor{Mao:2012:PSV} try to investigate possible influential factors such as the type of the instance, OS image size or VM location for different providers. The study provides researchers a basic understanding of the contributing factors which should be constant during their experiments and also mean average times to be taken into account in the simulation of cloud environment \cite{Grozev:2014:MPL}.

\subsubsection{Vertical Scaling}
While horizontal scaling is a conventional strategy in the management of resources offered by providers, the advent of virtualization techniques has introduced new resource level scaling opportunities including elastic VMs. In vertical scaling, the VM elasticity feature is utilized to change existing VM configurations, virtual cores or RAM, on the fly to adapt to new requirements of the system \citep{Yang:2013:WPA, Molto:2016:AMV}. The online VM reconfiguration without turning the VM off is getting more attention especially when the time and cost factors are considered in RMM decisions. First of all, maintaining SLA objectives becomes challenging when there are sudden changes and spikes in the workload, and RMM needs a quick solution to increase the resources in the system. In horizontal scaling, the time it takes to launch new VMs can be a bottleneck when a fast effective solution is needed. On the other hand, new VMs require new softwares to be installed which can lead to additional license costs. Elastic VMs can offer the required concepts and technologies to implement practicable strategies for these problems. Indeed, depending on the application, resource level scalings might be the most fitted idea in the case of resource shortages. For example, a CPU saturated system hosting a CPU intensive application can be scaled by adding a new core to the VM without wasting memory, bandwidth and other resources. Moreover, the same VM can continue the execution of existing requests without a need for interruption or transferring their execution profiles to a separate VM.

However, the elastic VMs need the support from both the guest VM OS and hypervisor. Therefore, due to added complexity, many providers such as Google and Amazon do not offer this functionality. To have a better understanding of the extent of support for vertical scaling, one can refer to work done by \cite{Turowski:2015:VSC} which studies this capability for some of the common hypervisors and guest OS as well as OpenStack framework in the cloud environment. \cite{Profitbricks} is one of the Infrastructure as a Service (IaaS) providers that supports the live vertical scaling of CPU cores, RAM, Network Interface Card and Hard disks.
It should be highlighted that the capabilities of vertical scaling for the VM is limited to the physical host. This means that the amount of resources which can be added to the VM can not be more than the available resource on the physical host. Therefore, in the case of resource shortage on the host, a new host with the available amount of resources can be selected and VM migrations precede vertical scaling to provide required resources for upgrading VM configuration. 

\subsection{Load Distribution}
In a large scalable environment, we can find many replicas of one service such as database or application server components. The problem of distributing application requests among existing replicas of one component is a functionality provided by load balancers. A load balancer such as haProxy can be configured with a weight for each replica and distribute the loads according to this configuration among multiple VMs. Amazon Web Service Elastic Load Balancer (AWS ELB) \cite{amazon:2018} offers a simple round-robin strategy which assigns an equal weight for the all active VMs and selects them from a list in the order of appearance. Therefore, in a round-robin based balanced environment, the utilization of all VMs is affected similarly. Noticeably, this approach does not provide a cost or energy efficient solution especially in an underutilized environment, where a small fraction of the provided computing and storage resources are enough to maintain the required SLA. A more efficient version of this approach assigns weights based on the specific server characteristics such as the load of the server which helps to send new requests to least utilized servers first \cite{Teo:2001:CLB}. \citeauthor{Grozev:2014:MPL} follow this approach to consolidate web requests in a few servers without violating the QoS. The proposed method monitors CPU and RAM utilizations of the servers along the availability of the network buffer capacity to assign new requests. \citeauthor{Soni:2014:NOLB} prioritize the servers based on their hardware characteristic, distributing loads based on the computation power and availability of the machines. 
These approaches help the system to balance the workload among existing resources to keep the performance at the acceptable level. However, load distribution is a task level resource management which does not control the amount of resources in the system. Therefore, they should be integrated with other scaling methods such as horizontal and vertical scaling and migrations which help to maintain enough resources to handle the existing workload of the application.
\subsection{Load Shedding}
Load shedding is a type of self-healing approaches, primarily used in electrical power management to handle high loads in the system. The idea behind this approach is to maintain the availability of the system by sacrificing some quality of service in the presence of faults. In data stream mining, load shedding refers to mechanisms that try to find when and how much of data can be discarded so the system can continue working with an acceptable degradation of performance \citep{Tatbul:2003:LSD}. This approach is not effective for the applications with the highly dynamic workload and strict QoS requirements \cite{Kleiminger:2011:BLS}. However, one can consider request admission and resource reservation policies in the system in terms of the SLA agreements so the extra requests that can not be handled by available resources will be rejected, saving time and money for both users and service providers.

\subsection{VM Migration}
The scalability feature of cloud systems is highly dependent on virtualization technology that enables multiple applications to reside on one physical machine by sharing available resources. VMs are preferably not dependent on one machine and can be moved among different hosts when needed. This capability brings new opportunities for improving the utilization of cloud applications by finding proper placement of VMs in the system \citep{Wood:2007:BGS}. Migration can be done to fulfill different objectives. Migrating VMs from underutilized hosts and consolidating them in few hosts enable the system to shut down unused hosts, saving costs and energy. \citeauthor{Duggan:2016:RLA} investigate this problem while considering the available bandwidth as a factor to determine the best time for migrations. An over-utilized host, where the guest VMs are consuming all the available resource offered by the host, is another target which can benefit from migration. In this case, the application on the overloaded host may experience performance degeneration as there are requests that can not be handled in time due to the saturated resource and long waiting queues. Accordingly, \citeauthor{Sommer:2016:PLB} propose a prediction based migration strategy to find the overloaded hosts and triggers migration procedure to move some of the VMs to existing underloaded hosts. A combination of multiple forecasting methods is employed to predict the future resource consumptions of the VMs and identify the possible overloaded hosts based on the predicted values.

Migration is also a complementary procedure when the host can not fulfill the requests to increase the resource capacity required for vertical scaling actions. PREPARE proposed by \citeauthor{Tan:2012:PPP} utilizes this strategy to correct the performance problems caused by internal faults or load anomalies. The live migration is called when the vertical scaling action is ineffective or not possible due to the lack of the resources on the host.
 
 \begin{landscape}
{\scriptsize

\begin{longtable}{ | P{0.5cm} | P{1.5cm} | P{2cm} | P{1cm} | P{3cm} | P{1.5cm} | P{1cm} | P{1.5cm} | P{2cm} | }
	
	\caption[f]{Comparison of Data Aware Performance Management Approaches in Large Scale Systems} \label{tab:title}\\
		\hline

		\toprule
		Work  & Data Level & Learning Approach (ML: Machine Learning, RL: Reinforcement Learning)& Anomaly Aware & Anomaly problem & Cause Inference Level & Proactive & Resource Adjustment Techniques (H: Horizontal, V:Vertical) & Module \\
		\midrule
		\cite{di:2014:google} & System & ML    & X    & \_    & \_    & \checkmark     & Load balancing    & Data \\
		\midrule
		\cite{cetinski:2015:ame} & System & ML, Statistical & X    & \_    & \_    & X     & \_    & Data \\
		\midrule
		\cite{Yang:2013:WPA} &  System & Threshold, ML & X     & \_    & \_    & \checkmark & V, H  & Data, Plan \\
		\midrule
		\cite{Iqbal:2011:ARP} & System, Application & Threshold, Statistical & X     & \_    & \_    & \checkmark  & H     & Data, Plan \\
		\midrule
		\cite{Cunha:2012:SPA} & System, Application & ML    & \checkmark     & \multicolumn{1}{c|}{} & \_    & X     & \_    & Data \\
		\midrule
		\cite{Ashfaq:2014:ITFS} & Network & Statistical & \checkmark    & Network & \_    & X & \_    & Data\\
		\midrule
		\cite{Nguyen:2013:fchain} & System, Structure & ML & \checkmark     & Software bug, Resource bottleneck & Component, Metrics & X     & \_    & Data \\
		\midrule
		\cite{Guan:2013:AAI} & System   & ML    & \checkmark     & Resource bottleneck & Type of Anomaly & X     & \_    & Data \\
		\midrule
		\cite{gu:2009:online} & System, Application & ML    & \checkmark    & Resource bottleneck & Type of Anomaly & \checkmark     & \_    & Data \\
		\midrule
		\cite{Cid:2015:OBI} & System   & ML    & \checkmark     & Deadlock, Starvation, Livelock & Type of Anomaly & \checkmark    & \_    & Data \\
		\midrule
		\cite{Dean:2012:UUB} & System   & ML    & \checkmark     & CPU/Mem leak, Network hog & Metrics &  \checkmark    & \_    & Data \\
		\midrule
		\cite{Tan:2012:PPP} & System   & ML    & \checkmark     & Resource bottleneck & Metrics & \checkmark     & V, Migration & Data, Plan \\
		
		\midrule
		\cite{Liu:2017:AHFPM} & System  & RL, ML & X     & \_    & \_    &  \checkmark  & H     & Data, Plan \\

		\midrule
		\cite{Hatef:2018:HIDCC} & Network  & Signature, ML & \checkmark    & Network   & Type of anomaly    &  \checkmark  & \_     & Data \\

		\midrule
		\cite{Wang:2016:AFD} & System, Application & ML, Statistical  & \checkmark     & Resource bottleneck & Metrics & X     & V, Migration & Data, Plan \\
		\midrule
		\cite{Xiong:2013:VAM} & System, Application & ML, Statistical & \checkmark     & Resource bottleneck & Metrics & X     & \_    & Data \\
		\midrule
		\cite{Nguyen:2011:PPR} & System   & Statistical & \checkmark     & Resource bottleneck, Offload bug, Load balancing bug & Component & X     & \_    & Data \\
		\midrule
		\cite{Dean:2014:PTR} & System & Statistical & \checkmark     & Resource bottleneck, software bugs, multi-tenancy problem, network packet loss, deadlock & Type of Anomaly(external vs internal) & X     & \_    & Data \\
		\midrule
		\cite{Dean:2014:PPO} & System & ML, Signature & \checkmark     & Software bugs & Code level & X     & \_    & Data \\
		\midrule
		\cite{Tucek:2007:TDP} & Application & Statistical  & \checkmark     & Software bugs & Code level& X     & \_    & Data \\
		\midrule
		\cite{Zhang:2016:TI} & System   & ML    & \multicolumn{1}{c|}{} & Resource bottleneck, Database abuse & Target Thread/Process & X     & \_    & Data \\
		\midrule
		\cite{Islam:2012:EPM} & System   & ML, Statistical & X     & \_    & \_    &  \checkmark   & \_    & Data \\
		\midrule
		\cite{ Shamsolmoali:2014:SFS} & Network   & Statistical  & \checkmark     & DDos Attacks & \_    & X     & \_    & Data \\
		\midrule
		\cite{ Cohen:2004:CID} & System, Application & Statistical  & \checkmark     & Resource bottleneck & Metric & X     & \_    & Data \\
		\midrule
		\cite{MI:AAPC:2008} & System, Application & Signature & \checkmark     & \_    & \_    & X     & \_    & Data \\
		\midrule
		\cite{Cherkasova:2009:AAD} & System, Application & Statistical, ML, Signature & \checkmark     & Resource bottleneck, Application update & \_    & X     & \_    & Data \\
		\midrule
		\cite{BRUNNERT:CPECPU:2017} & System, Application, Structure & Signature, Statistical & X     & \_    & \_    & X     & \_    & Data \\
		\midrule
		\cite{Sharma:2013:CPD} & System, Application & Statistical, Signature, ML & \checkmark     & Load, Software bugs  & Metrics, Type of Anomaly & X    & V, H, Migration & Data, Plan \\
		\midrule
		\cite{Gmach:2009:RPM} & System   & Threshold & \_    & \_    & \_    & \checkmark  & H, Migrations & Data, Plan \\
		\midrule
		\cite{Han:2012:LRS} & System, Application & Threshold & X & \_    & \_    & X     & H, V  & Data, Plan \\
		\midrule
		\cite{Hong:2015:DDA} & System   & ML    & \checkmark     & Resource bottleneck & \_    & X     & \_    & Data \\
		\midrule
		\cite{Grimaldi:2015:FCA} & System   & Control Theory & \checkmark     & Load, Hardware failure & \_    & X & H     & Data, Plan \\
		\midrule
		\cite{Kalyvianaki:2009:SSC} & System, Application & Control Theory &X & \_    & \_    & X     & V & Data+Plan \\
		\midrule
		\cite{Al-Shishtawy:2013:EEM} & System, Application & control Theory & X & \_    & \_    & X     & H     & Data+Plan \\
		\midrule
		\cite{Wang:2011:SAD} & System & Statistical  & \checkmark     & Resource bottleneck, Software bugs  & \_    & X     & \_    & Data \\
		\midrule
		\cite{Ibidunmoye:2017:BDS} & System   & Statistical  & \checkmark    & Resource bottleneck & \_    &  X & \_    & Data \\
		\midrule
		\cite{Peiris:2014:PAD} & Application, System & Threshold, Statistical & \checkmark     & \_    & Metrics & X    & \_    & Data \\
		\midrule
		\cite{Matsuki:2016:RCA} & System, Application & Threshold, Statistical & \checkmark     & Resource bottleneck & Metrics & X     & \_    & Data \\
		\midrule
		\cite{Wajahat:2016:MLA} & System, Application & ML, Statistical & X     & \_    & \_    &  \checkmark    & H     & Data, Plan \\
		\midrule
		\cite{Gu:2005:DAN} & Network   & Statistical  & \checkmark  & Port scan & Packet Information & X    & \_    & Data \\
		\midrule
		\cite{Cao:2014:EDS} & System   & Statistical & \checkmark     & DoS Attack & \_    & X    & \_    & Data \\
		\midrule
		\cite{Ajila:2016:MLCP} & System, Application & ML    & X     & \_    & \_    & \checkmark     & \_    & Data \\
		\midrule
		\cite{Li:2015:SCVM} & System   & ML    & \checkmark     & Resource bottleneck & \_    & \checkmark     & VM Placement & Data, Plan \\
		\midrule
		\cite{Dutreilh:2010:DCR} & System, Application & Threshold, RL & X & \_    & \_    & X     & H     & Data, Plan \\
		\midrule
		\cite{Yazdanov:2013:VAV} & System   & RL, Statistical & X   & \_    & \_    & \checkmark   & V     & Data, Plan \\
		\midrule
		\cite{Duggan:2016:RLA} & System   & RL    & X     & \_    & \_    & X     & Migration & Data, Plan \\
		\midrule
		\cite{Arabnejad:2017:CRL} & System, Application & RL  & X & \_    & \_    & X     & H     & Data, Plan \\
		\midrule
		\citep{Qu:2016:RCA} & System   & Signature   & \checkmark     & Resource Shortage & \_    & X     & H, Over-provisioning & Data, Plan \\
						
		\midrule
		\cite{Shen:2011:CER} & System, Application  & Signature,  ML &  \checkmark     & Resource bottleneck    & \_    & \checkmark    & V, Migration     & Data, Plan \\
				
		\midrule
		\cite{Molto:2016:AMV} & System   & Threshold & \checkmark    & Resource bottleneck & \_    & X     & V, Migration & Data, Plan \\
		\midrule
		\citep{Wood:2007:BGS} & System, Application & Threshold, Statistical & \checkmark     & Resource bottleneck & \_    & \checkmark  & Migration & Data, Plan \\
		\midrule
		\cite{Sommer:2016:PLB} & System   & Statistical & X     & \_    & \_    & \checkmark     & Migration & Data, Plan \\
		\midrule
		\cite{Zhuang:2014:NND} & System   & Rule  & \checkmark     & Network related Application bugs & Target System Calls &  X    & \_    & Data \\
		\midrule
		\cite{amazon:2018} & System   & Threshold & X     & \_    & \_    & R     & H     & Data, Plan \\

		\bottomrule

\end{longtable}

}
\end{landscape}

\section{Gap Analysis and Future directions}\label{Gaps}
Based on the analysis of different aspects of performance-aware resource management in clouds, a number of challenges have been observed that needs to be investigated more thoroughly. Accordingly, we propose a list of the potential research areas and future directions in the following sections.

\subsection{VM Elasticity Analysis }
Vertical elasticity is one of the rather new functionalities introduced for cloud resource scaling which has not yet been prevalent compared to horizontal scaling. Technical limitations to support VM elasticity along the higher complexity of fine-grained scaling (CPU and RAM level compared to VM level) management in data centers makes this approach limited in practice and most known public providers do not offer this service. Therefore, a more detailed study of the characteristics of vertical elasticity especially time sensitivity analysis for different resource types, OS and hypervisors is required to help the design of accurate solutions in the area of proactive resource management.

\subsection{Dynamic Algorithms in Cloud}
The effectiveness of many advanced machine learning methods is highly dependent on pre-configurations that define a proper threshold or parameter values, usually based on the characteristics of data and applications. There are many works targeting the problem of parameter configuration of learning methods in the general field of data prediction and anomaly detection. However, considering the dynamic nature of cloud environment and hosted applications, applying the traditional protocols for initialization of these algorithms may not deliver the best of their potential. Having an automated procedure for configuration and dynamic threshold settings are important to be further investigated.

\subsection{Anomaly Cause Inference}
Anomaly cause inference has been studied as part of the data analysis module to help RMM better decide on selecting corrective actions. However, existing works in this area mostly are coarse-grained, giving suggestions about the bottleneck metrics based on the resource level information. Moreover, the contribution of knowledge from cause identification in the process of RMM planning and resource management has not been fully investigated. A more autonomic and integrated cause identification process which has continuous interaction with planning module to get timely feedbacks on the quality of information is a challenge for future researchers.

\subsection{Realistic View for Anomaly Aware Auto-Scaling}
In this paper, we have tried to cover both areas of data analysis including prediction and anomaly detection as well as the resource adjustment planning for performance management. The significance of having timely performance alerts ahead of time for planning module to have enough time for preparation and triggering actions makes the real-time sensitivity analysis a challenge which can be only investigated in a real environment. There are several technical limitations that can affect the performance of both parts which can only be detected during a real implementation of the complete framework including all interactions among distribute components.

\subsection{Application Dependent Detection Accuracy Trade-offs}
In data analysis part of the investigated problem, Area Under the Curve (AUC) commonly is used to show the performance of anomaly detection algorithms in detecting anomaly points. However, in the area of cloud performance analysis, the collected data usually has a characteristic that the number of normal instances are much higher than anomaly points. The lack of the balance in the number of instances for different classes raises the question of whether AUC metric is biased by true negative points. We believe that presenting the performance results by comparing both metrics AUC and Precision-Recall Area Under the Curve (PRAUC) which demonstrate the functionality of the algorithms from different points of the view is an important part of the anomaly detection problems in this area. 
This is a point that can be very important for some applications which require complex recovery points in the case of the true anomaly events. For example, for prevention mechanisms that target disk related problems with expensive mitigation actions, a solution with higher precision and minimum of false alarms may be preferred. Referring to our survey, this is an interesting point which is highly neglected and requires a detailed analysis of the effectiveness of proposed anomaly detection methods considering service owner preferences.

\section{Conclusions}\label{Summary}
With the emergence of cloud computing and the power of data analytic methods, new opportunities for improvement of performance-aware distributed resource management mechanism has been introduced. Analysis of collected data from measurable system attributes gives valuable information about the health state of the system and the performance of the hosted applications.

In this paper, we investigate different approaches in the performance management of cloud environment. Identifying the major limitations and considerations in the selection of the best strategy for proper resource configuration highlights the need for more complex and automate procedures to handle the dynamism of the environment. We have proposed a taxonomy of problem focusing on the value of the data as a source of knowledge for resource management decision making and presented a survey of the existing works in the field of performance-aware cloud resource management accordingly. The listed categories in the taxonomy are defined based on the characteristics of the reviewed works including presented architecture, the granularity of collected performance data, targeted performance problem and the types of resource management actions. Based on the reviewed works, a list of observed gaps and possible directions is suggested which can give new insights and starting points for future researchers.

\bibliographystyle{ACM-Reference-Format}
\bibliography{ref}


\end{document}